\documentclass[amsmath,twocolumn]{revtex4}

\usepackage{graphicx} 
\usepackage{dcolumn} 
\usepackage{bm} 
\usepackage{hyperref}

\begin{document} 

\title{Precision polarimetry with real-time mitigation of optical-window
  birefringence}

\author{B. K. Park}
\email{novakyu@ocf.berkeley.edu}
\author{A. O. Sushkov}
\affiliation{Department of Physics, University of California, Berkeley,
  California 94720-7300}
\author{D. Budker}
\affiliation{Department of Physics, University of California, Berkeley,
  California 94720-7300 \\ and Nuclear Science Division, Lawrence
  Berkeley National Laboratory, Berkeley, California 94720}

\date{January 31, 2008}

\begin{abstract}
Optical-window birefringence is frequently a major obstacle in
experiments measuring changes in the polarization state of light
traversing a sample under investigation.  It can contribute a signal
indistinguishable from that due to the sample and complicate the
analysis.  Here, we explore a method to measure and compensate for the
birefringence of an optical window using the reflection from the last
optical surface before the sample.  We demonstrate that this arrangement
can cancel out false signals due to the optical-window
birefringence-induced ellipticity drift to about 1\%, for the values of
total ellipticity less than 0.25 rad.\\ 
(DOI: \href{http://dx.doi.org/10.1063/1.2835902}{10.1063/1.2835902})
\end{abstract}

\maketitle

\section{Introduction} 
Kerr effect, measured in liquid helium~\cite{liquidhe} and liquid
nitrogen~\cite{liquidn}, is a useful method for non-contact measurement
and monitoring of the electric fields in low-temperature experiments.
In particular, we have proposed to use this technique to monitor the
electric field set-up in the new search for the neutron electric dipole
moment~\cite{nedm}.  However, stress-induced birefringence of optical
windows puts a limit on the sensitivity of this method.  While constant
stress due to the window mounting can be taken into account in modeling
the signal by prior measurement of the optical-window
birefringence~\cite{unianalysis,ellipswindow}, or optically compensated
in the experimental setup by use of retarder
compensators~\cite{optcompone,optcomptwo,windowdesign}, for experiments
requiring greater sensitivity, the time-dependent stress due to
temperature drifts cannot be treated using the same methods, because the
amount of retardation due to the optical window at a given moment is not
known.

A careful experimental design, such as using low-stress window mounts
and controlling the window temperature, can minimize this
birefringence-induced polarization drift to as little as
$\ensuremath{3.5 \times 10^{-5}~\mathrm{rad}}$ over the measurement time
of 12 h~\cite{lowdrift}.  But this is still not good enough for more
sensitive experiments, such as measurement of Kerr effect in liquid
helium.  In a practical experimental setup, we expect Kerr effect
signals of the order of $10^{-5} - 10^{-6}$ rad~\cite{liquidhe}.  In our
laboratory, where low-temperature optical access is implemented with a
set of 1-in.-diameter, 2-mm-thick, epoxy-mounted fused-silica windows,
testing with 632 nm light showed that window birefringence introduces
offsets in polarization rotation and ellipticity, the latter being
defined as inverse tangent of ratio of the axes of the polarization
ellipse, on the order of 0.1 rad, which also depend on the window
temperature. This offset drifts within a range of about $10^{-4}$ rad
over periods of hundreds of seconds, and this noise overwhelms the
typical signal from the Kerr effect of liquid helium.

In past experiments, with the exception of those where the residual
birefringence in instruments is small enough to be
ignored~\cite{kerrliquids}, various techniques have been used to
distinguish the effects of the sample from those of the optical-window
birefringence.  In experiments with an isotropic sample, the incoming
polarization of light can be rotated to distinguish the anisotropic
feature of optical-window birefringence~\cite{jellison}.  In other
experiments, notably Kerr-effect experiments, modulation of the applied
electric field is used to distinguish the signal from background
noise~\cite{liquidhe}.  Neither of these methods can be used in the
application of Kerr effect for monitoring electric fields. Thus, we
explore the feasibility of measuring and compensating for the
optical-window birefringence drifts using the reflected light from the
last optical surface before the sample.

\section{Experimental setup} 

\begin{figure}
\includegraphics[width=0.45\textwidth]{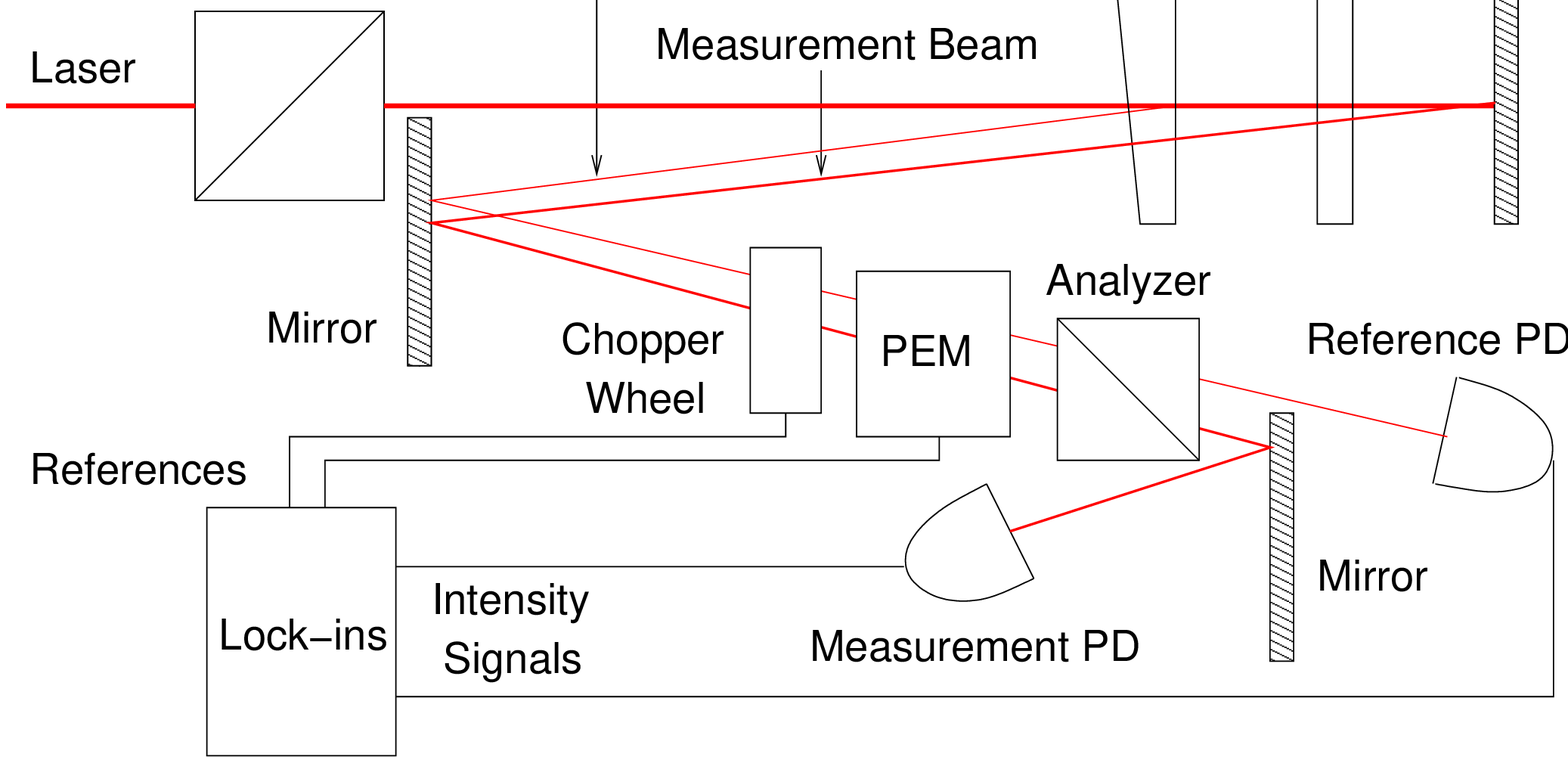}
\caption{\label{fig_1}Schematic diagram for the setup}
\end{figure}

A typical experimental situation is simulated with two fused-silica
parallelepipeds squeezed along the (approximately vertical) axis, making
an approximate angle of $\pi/4$ with the azimuth of the incident light
polarization.  These are labeled as ``optical window'' and ``sample'' in
the schematic diagram of our experimental setup shown in
Fig. \ref{fig_1}.  In the absence of stress on the optical window, the
maximum ellipticity that can be introduced by the sample on a single
pass (depending on orientation) is equal to half of the
birefringence-induced phase difference, given by, $\phi = 2 \pi d (n_s -
n_f) / \lambda$, where $d$ is the thickness of the sample, $\lambda$ is
the wavelength of the probe beam, and $n_s$ and $n_f$ are the indices of
refraction along the slow and fast axes, respectively.  We place a
mirror just after the sample, so that the light double-passes the
sample, and the entrance window is also the exit window.  This
arrangement works for measuring the Kerr effect since the effect of a
small birefringence is additive for the double pass.  The sample
birefringence is what we want to measure, but this signal is obscured by
the birefringence of the window.  To mitigate this problem, we measure
the window birefringence using the reflection from the back surface of
the window.

The laser light at 632 nm with intensity of 1 mW is produced by a
Hitachi laser diode HL6320G mounted on a Thorlabs TCLDM9 mount, with
laser current and temperature controlled by Thorlabs LDC 201ULN and TED
200 controllers, respectively.  The light is collimated to a beam width
of about 1 mm diameter and is linearly polarized at $\pi/4$ rad from the
optical axis of the sample by passing through a Glan-Thompson polarizer.
The beam is reflected back by a near-normal-incidence dielectric mirror
placed after the sample to pass through the optical-window and the
sample twice.  Then it goes through a fused-silica photoelastic
modulator (PEM) operating at 50 kHz, whose optical axis is aligned
vertically, and a Wollaston-prism analyzer, which makes an angle of
$\pi/4$ rad with the optical axis of the PEM and, in this particular
setup, is crossed with the Glan-Thompson polarizer.  Additional
dielectric mirrors were placed to ensure that all angles of reflection
before the analyzer are less than 5$^\circ$ and to overcome space
limitation for the modulation polarimeter.  For angles of incidence up
to 20$^\circ$, a commercial polarimeter, Thorlabs PAX5710VIS, was used
to check that the dielectric mirrors used here do not alter the
polarization state within the sensitivity of 1 mrad for ellipticity and
0.1 mrad for polarization rotation.  For lock-in detection, and for
calibration purposes, we introduce light modulation at 700 Hz with a
chopper wheel. The measurement and the reference beams are calibrated
separately (as explained below) using the synchronously detected signals
from each photodiode.  All data are collected with a LabVIEW program
through the General Purpose Interface Bus (GPIB).

The main experimental limitation is posed by the fact that the window
birefringence is nonuniform. This can impart different ellipticity to
the measurement beam and the reference beam (Fig. \ref{fig_1}).  To
minimize the effect of this nonuniformity, we superimpose the two beams
as they pass through the window. In order to be able to separate the
measurement and reference beams, as well as the reflection from the
front surface of the window, a wedged window is used (a wedge angle of
$1^\circ$ is sufficient in our setup).  In our setup, we were able to
reduce the distance between the two beams as they pass through the
window to roughly one-tenth of the $\approx 1$ mm beam diameter.

The photodiode signal with the modulation polarimeter setup is given by,
\begin{align}
I(t) &= \frac{K I_0}{2}\{1 - \cos(2\epsilon) \sin (2 \alpha) \cos(A \sin
\omega t) \nonumber \\
&\hspace{5mm} + \sin(2 \epsilon) \sin(A \sin \omega t)\}, \label{photosignal}
\end{align}
where $\epsilon$ and $\alpha$ are the polarization ellipticity and
azimuth of the light incident on the PEM, $I_0$ is the intensity after
the polarizer, before the window and the sample, and $K$ represents the
reflection and absorption losses by the optical components before the
photodiode.  With $\alpha \approx \pi/4$ if no additional rotation is
introduced by the sample and the optical window, $A$ is the peak
differential phase retardation introduced by the PEM, and $\omega
\approx 2\pi \times \ensuremath{50~\mathrm{kHz}}$ is the PEM modulation
frequency. The sine and cosine terms can be expanded in terms of Bessel
functions of the first kind, multiplied by odd and even harmonics of
$\omega$, respectively. In the present experiment, we detect the dc
signal and the first harmonic signal. These two signals are used for
sensitive detection and calibration of $\epsilon$~\cite{pem}.
\begin{align}
I(t) &= \frac{K I_0}{2}\{1 - \cos(2\epsilon) \sin (2 \alpha) J_0(A)
\nonumber\\
&\hspace{5mm} + 2\sin(2 \epsilon) J_1(A) \sin(\omega t) + \cdots
\}.\label{workingEq}
\end{align}
Because the first-harmonic term depends only on $\epsilon$ and is
completely insensitive to $\alpha$, this is a good polarimeter setup for
measuring the light ellipticity.  Furthermore, because 2.405 rad is a
zero of the zeroth-order Bessel function of the first kind
($J_0(\ensuremath{2.405~\mathrm{rad}}) \approx 0$), when the optical
axis of the PEM forms the angle of $\pi/4$ rad with the analyzer,
setting $A = \ensuremath{2.405~\mathrm{rad}}$ ensures that the dc offset
is independent of incoming light-polarization~\cite{pem}. This fact can
be used to check and calibrate the PEM modulation amplitude to 2.405 rad
by, for example, placing a wave plate before the PEM and rotating it.
PEM modulation amplitude is varied until the dc offset does not change
with the rotation of the wave plate. In our setup, this method was used
to calibrate the PEM retardation with approximately 1\% accuracy, and
the variation of PEM amplitude within the separation between the two
beams as they traverse the PEM is within this accuracy. For this special
value of $A = 2.405~\mathrm{rad}$, Eq. (\ref{workingEq}) simplifies to
\begin{equation}
I(t) = \frac{K I_0}{2}\{1 + 2 J_1(A) \sin(2 \epsilon) \sin(\omega t)\}.
\end{equation}

The modulation polarimeter is thus calibrated for measuring ellipticity
by dividing the intensity detected at the first harmonic of the PEM
operating frequency, $\omega$, by the dc offset. Thus, $\epsilon$ is
given by
\begin{equation}
\label{signal}
\epsilon = \frac{1}{2}\arcsin\left(\frac{I_1}{2 I_{dc} J_1(A)}\right),
\end{equation}
where $A = \ensuremath{2.405~\mathrm{rad}}$, and $I_{1}$ and $I_{dc}$
are the intensity at the first harmonic of PEM operating frequency,
measured with a Signal Recovery 7265 lock-in amplifier, and the DC
offset, measured at 700 Hz with a second Signal Recovery 7265 lock-in
amplifier referenced to the chopper wheel, respectively.  Before taking
the ratio of two signals thus measured, we multiply the 50 kHz signal by
an additional factor of 2 to account for reduction in intensity by 700
Hz chopping.  Because the intensity signal itself contains the
calibration information, this calibration can be done for both beams
separately and, with additional lock-in amplifiers,
simultaneously. Also, since $I_1$ and $I_{dc}$ both depend linearly on
$K I_0$, this calibration is independent of laser intensity fluctuations
or reflection losses in optical components.  For large values of
ellipticity it was checked that this calibration method yields an
agreement within 5\% with Thorlabs PAX5710VIS, but the Thorlabs
polarimeter cannot be reliably used to measure ellipticities smaller
than 0.01 rad.

\section{Jones-matrix analysis}

At nearly-normal incidence, up to leading order in the angle of
reflection $\theta_r$, reflection coefficients for light polarized
parallel and perpendicular to the reflection plane are, respectively,
\begin{align}
\label{Rpara} 
R_{\parallel} &= \frac{(n_1-n_2)^2}{(n_1+n_2)^2}\left(1 +
2\frac{n_2}{n_1} \theta_r^2\right), \\
\label{Rperp} 
R_{\perp} &= \frac{(n_1 - n_2)^2}{(n_1 + n_2)^2} \left(1 - 2
\frac{n_2}{n_1}\theta_r^2\right).
\end{align}
This gives about $5\%$ reflection at each optical surface for typical
optical-window materials, in the absense of inteference effects from
highly parallel surfaces. Also, at a nearly-normal incidence, to first
order in the angle of incidence, light is reflected without change in
the polarization state.

For the experimental setup described above (Fig. \ref{fig_1}), in the
Jones-matrix formalism, the polarization states for the measurement beam
($\psi_m$) and the reference beam ($\psi_r$) are given by
\begin{align}
\psi_m &= M_w M_s M_m M_s M_w \psi_0, \\
\psi_r &= M_w M_m' M_w \psi_0, 
\end{align}
where, $\psi_0$ is the Jones vector for the initial polarization state,
and $M_w$, $M_s$, $M_m$, and $M_m'$ are the Jones matrices modeling the
birefringent optical-window, the sample, the dielectric mirror, and
reflection at the last optical surface of the window, respectively.  In
our setup using a dielectric mirror, $M_m$ is equal to identity as it
introduces negligible ellipticity and rotation. Likewise, at
nearly-normal incidence, $M_m'$ is equal to identity (except for
changing overall light intensity) as well, since there is no measurable
change in the polarization state upon reflection.  And $M_s$ and $M_w$
are given as follows, up to a constant phase factor:
\begin{align}
M_s &= \left(\begin{matrix} 
  e^{i \phi_s/2} & 0 \\ 
  0 & e^{-i\phi_s/2} \end{matrix} \right), \\
M_w &= U(\theta) \cdot \left(\begin{matrix} 
  e^{i \phi_w /2} & 0 \\ 
  0 & e^{-i \phi_w /2} \end{matrix} \right) \cdot U^{-1}(\theta),
\end{align}
where $\phi_s$ and $\phi_w$ are the birefringences of the sample and the
window, respectively, $\theta$ is the angle between the fast axes of the
birefringent window and the sample, and $U(\theta)$ is the rotation
matrix. We assume that the sample birefringence has a fixed alignment,
since it is controlled by the experimental setup.

In the special case where the fast axis of the birefringent window is
aligned with fast axis of the sample, the birefringences simply add.
The general case with misaligned optical axes presents a more difficult
problem for general values of $\phi_s$ and $\phi_w$~\cite{transpwave}.
However, as long as $\phi_s$ and $\phi_w$ are small, for $\psi_0$
linearly polarized at $\pi/4$ rad to the fast axis of the sample, up to
the terms linear in $\phi_s$ and $\phi_w$, we obtain
\begin{align} 
\label{psm} 
\psi_m &= \frac{1}{\sqrt{2}}\left(\begin{matrix} 
  1 - i\{\phi_s + \phi_w (\cos 2\theta + \sin 2\theta)\} \\ 
  1 + i \{\phi_s + \phi_w (\cos2\theta - \sin 2\theta)\}
\end{matrix}\right),  \\
\label{psr} 
\psi_r &= \frac{1}{\sqrt{2}}\left(\begin{matrix} 
  1 - i\{\phi_w (\cos 2\theta + \sin 2\theta)\} \\ 
  1 + i \{\phi_w (\cos2\theta - \sin 2\theta)\}
\end{matrix}\right).
\end{align} 
In terms of Jones vector components, the light ellipticity is given
by~\cite{ellipticity}
\begin{equation}
\epsilon = \frac{1}{2}\arcsin\left\{\frac{i(\psi_x^*\psi_y -
  \psi_x\psi_y^*)}{|\psi_x|^2 + |\psi_y|^2}\right\},
\end{equation}
where $\psi_x$ and $\psi_y$ are the complex upper and lower components
of a Jones vector $\psi$ respectively. To first order in $\phi_s$ and
$\phi_w$, the ellipticities at the output are
\begin{align}
\epsilon_m &= \phi_s + \phi_w \cos2\theta, \\
\epsilon_r &= \phi_w \cos2\theta.
\end{align}

The optical-window term appearing in the measurement beam can be
independently determined from the light reflected from the back surface
of the window. Thus, to first order in $\phi_s$ and $\phi_w$, we can
subtract the ellipticities of the two beams,
\begin{equation}
\epsilon_m - \epsilon_r = \phi_s.
\end{equation}

In the preceding calculations, if we retain higher-order terms, assuming
$\phi_s~\ll~\phi_w~\ll~1$, the leading error term is $\frac{1}{4} \phi_s
\phi_w^2 (1 - \cos 4\theta)$ for arbitrary orientation of optical-window
birefringence. For example, in an experimental setup measuring sample
birefringence on the order of $10^{-6}$ rad, with the optical-window
birefringence on the order of $0.1$ rad, this would lead to error of
less than $5 \times 10^{-9}$ rad, or a fractional error less than 0.5\%
of the sample birefringence. In contrast, for the same experimental
setup, but without the active cancellation of optical-window
birefringence, the error would be on the order $0.1$ rad and would
overwhelm the sample signal.

Additional considerations that may need to be taken into account are:
\begin{itemize} 
\item Rotation of polarization state of the reference beam introduced
  upon reflection at the back surface of the window: Since reflection
  coefficients as given in Eqs. (\ref{Rpara}) and (\ref{Rperp}) depend
  on the angle of reflection and indices of refraction, different
  fractions of incoming light are reflected for light polarization
  parallel and orthogonal to the fast axis of the medium, resulting in
  rotation.  For an order of magnitude estimate of this rotation, let
  the fast axis of the medium be parallel to the plane of reflection.
  Let $n_1 = 1$ and $n_2 = n_f$ for the light polarized along the fast
  axis, and $n_2 = n_s$ for the slow axis, and let $n$ be average of
  $n_f$ and $n_s$. In the limit where $(n_s - n_f) \ll n^2 - 1$ and
  $\theta_r \ll 1$, when the incident light is polarized at $\pi/4$, the
  resulting rotation is,
  \begin{equation}
    \Delta \alpha = \frac{(n_s-n_f)}{n^2-1} - n \theta_r^2.
  \end{equation}
  The quantity $(n_s - n_f)$ is related to the single-pass phase shift $\phi_w$ by:
  \begin{equation}
    n_s - n_f = (\lambda \phi_w)/(2 \pi d).
  \end{equation}

  Since our modulation polarimeter setup is insensitive to rotation, we
  consider only the change in the ellipticity due to the rotation. Given
  a small rotation $\Delta \alpha$ from $\pi/4$, to the leading order in
  $\phi_w$ and $\Delta \alpha$, the change in the ellipticity of the
  reference beam traversing the window on the return trip after
  reflection is,
  \begin{equation}
    \epsilon = \left(\frac{1}{2}-\Delta \alpha^2\right) \phi_w.
  \end{equation}

  When the typical values ($\lambda = \ensuremath{632~\mathrm{nm}}$, $d
  = \ensuremath{1~\mathrm{cm}}$, $n = 1.5$, and $\theta_r = 0.01$ rad)
  are plugged in, the fractional change of $\epsilon$ due to $\Delta
  \alpha$ is several orders of magnitude less than unity.

\item Ellipticity introduced into the measurement beam by the mirror: In
  the case of a dielectric mirror, residual stress in each dielectric
  layer can introduce additional ellipticity.  At room temperature, a
  separate measurement shows that the dielectric mirror used introduces
  ellipticity less than 1 mrad at incidences within $20^\circ$.

\item Spatial nonuniformity in the birefringence of window: If the
  birefringence in the window is not uniform and the two beams pass
  through different parts of the window on the return trip, they may
  acquire different ellipticity, a systematic indistinguishable from a
  genuine signal.  In our experiments this is the dominant source of
  error with 1\%--2\% change in birefringence per 1 mm.
\end{itemize}

\begin{figure*}
\includegraphics[width=\textwidth]{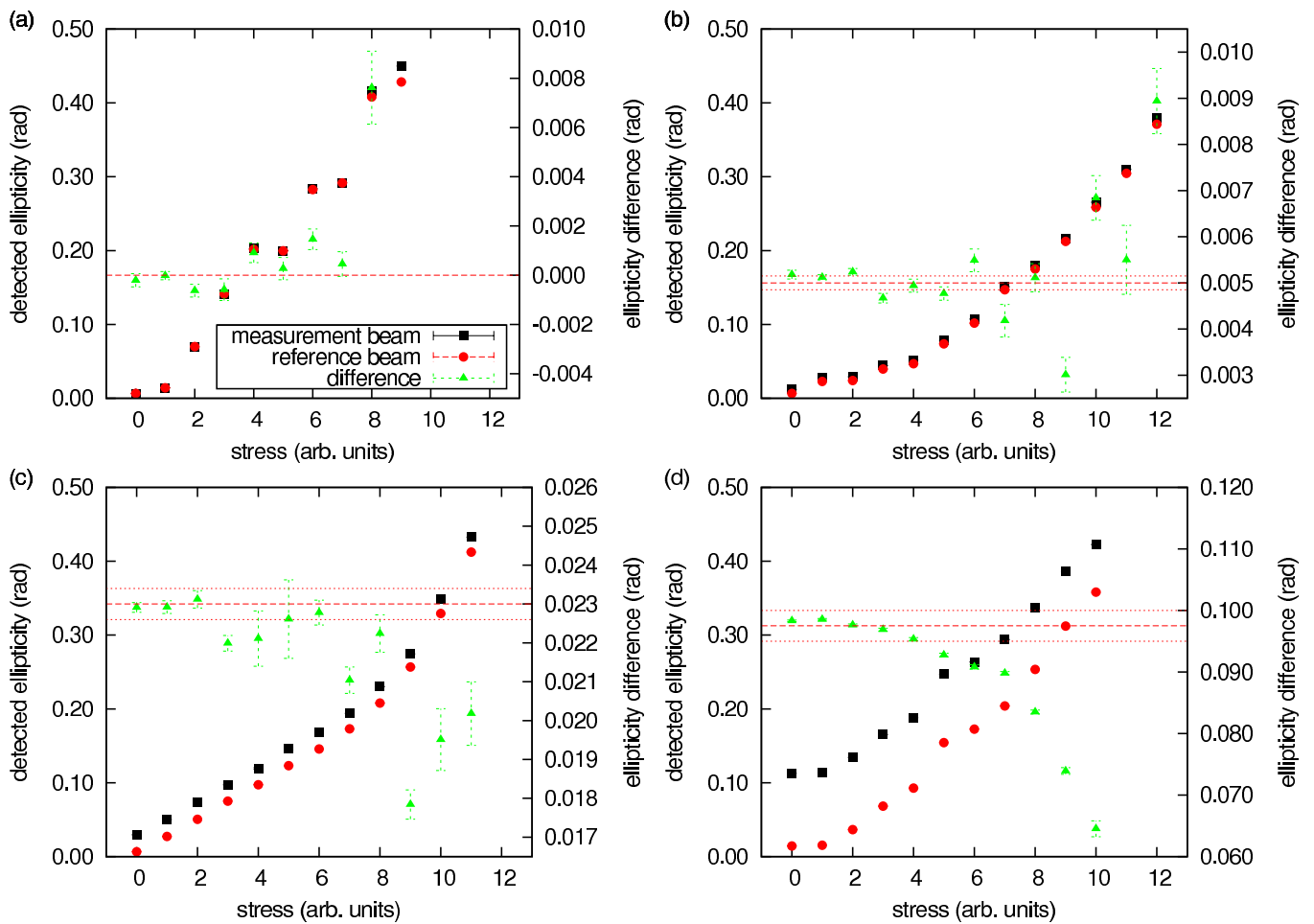}
\caption{\footnotesize Experimental results on cancellation of window
  birefringence at various sample ellipticities. The sample
  ellipticities are as represented by the dashed line, with the scale on
  the right. The dotted lines indicate the uncertainty in determination
  of the sample birefringence-induced ellipticity change.  The
  horizontal axis represents an increasing level of the stress on the
  window, and the left vertical axis shows the detected ellipticity. The
  right axis is for the difference of the ellipticity measured for the
  two beams. The error bars represent statistical errors and are
  obtained from the spread of the data for the measurement time of 10
  s.}
\label{plots}
\end{figure*}

\section{Results} 

The empty polarimeter has noise of $\ensuremath{1 \times
  10^{-5}~\mathrm{rad/\sqrt{Hz}}}$, but with either the window or the
sample in place, the temperature-dependent drift in the birefringence
dominates the error. Although in this experiment the ellipticity
measurements of the two beams were separated in time by about half a
minute, leading to imperfect cancellation, in principle, the drift in
window birefringence can be canceled out with simultaneous monitoring of
both beams.

As a check of the experimental setup, a series of measurements was made
without anything in the sample space.  The window birefringence is
varied by applying pressure on the fused-silica 1$^\circ$ optical wedge
with a vise.  In this case, we expect $\epsilon_s$ and $\epsilon_w$, the
ellipticities of measurement beam and the reference beam, respectively,
to be equal. The obtained results (Fig. \ref{plots}a) indicate that the
cancellation of the window birefringence is, indeed, possible, and allow
an estimate of the level to which such cancellation can be done.  We
obtain a result consistent with zero difference between the two signals,
and the spread of the points indicates that we can cancel out the window
birefringence to about $1\%$ of its value with the current setup,
limited by the nonuniformity of window birefringence and the finite beam
width.

For measurements with a sample, we take the sample birefringence to be
the difference in ellipticity detected in the measurement beam when the
sample is inserted and taken out at zero applied stress on the
optical-window. This difference is measured at the beginning and end of
each measurement, and the average is indicated by the dashed line. The
two measurements also give an order-of-magnitude estimate of the
uncertainty of the sample birefringence due to temperature-dependent
drifts, indicated by dotted lines on the plots. For small total
ellipticities, this estimate is comparable to the error of each data
point obtained from the spread of points over a 10-s measurement period.

The results for the measurement beam and the reference beam as the
stress on the optical-window is varied show a good agreement with the
measured sample ellipticity, particularly for small values of $\phi_s$
and $\phi_w$. At larger values of stress on the window, the
nonuniformity of window birefringence dominates the difference and
accurate measurement of the sample becomes more difficult. The
nonuniformity of optical-window birefringence is measured to be about
1\% change of total birefringence over the beam width of 1 mm for our
setup.

At larger values of $\phi_s$ and $\phi_w$, we see the second-order
effects due to the rotation introduced by the birefringent window. In
principle, these effects can be compensated for in the analysis
numerically using Jones-matrix formalism, provided that the azimuth
$\alpha$ is measured in addition to the ellipticity.  Since, as shown in
Eq. (\ref{photosignal}), the even harmonics of $\omega$ are sensitive to
$\alpha$, the rotation can be measured by using a lock-in amplifier to
recover the second-harmonic amplitude.

\section{Conclusions}

This work demonstrates that the birefringence of a window can be
measured using the reflection from the last optical surface before the
sample. This measurement can be done simultaneously with the measurement
of the total birefringence of the optical-window and the sample,
allowing for a real-time correction for window birefringence drifts
during the experiment. This cancellation can effectively be done to the
precision of $1\%$ of total ellipticity drift, limited by the spatial
optical-window birefringence nonuniformity, for the values of
optical-window birefringence less than 0.25 rad.

This work has been supported in part under Los Alamos National
Laboratory Project 20040104DR, ``Testing Time-Reversal Symmetry with
Ultracold Neutrons and with Solid State Systems,'' and by the National
Science Foundation through grant \#0554813.

\end{document}